\documentclass[aps,prl,two column]{revtex4}%
\usepackage{graphicx}
\usepackage[pdftex]{color}
\usepackage{amsmath}
\usepackage{amsfonts}
\usepackage{amssymb}%

\providecommand{\U}[1]{\protect\rule{.1in}{.1in}}

\def \be{\begin{equation}}
\def \ee{\end{equation}}
\def \bea{\begin{eqnarray}}
\def \eea{\end{eqnarray}}

\begin{document}
\title{Countering a fundamental law of attraction with quantum wavepacket engineering}
\author{G. Amit}
\affiliation{Faculty of Engineering and the Institute of Nanotechnology and Advanced Materials, Bar Ilan University, Ramat Gan 5290002, Israel}
\affiliation{Soreq Nuclear Research Center, Yavne, Israel}
\author{Y. Japha}
\affiliation{Department of Physics, Ben-Gurion University of the Negev, Be’er Sheva 84105, Israel}
\author{T. Shushi}
\affiliation{Department of Physics, Ben-Gurion University of the Negev, Be’er Sheva 84105, Israel}
\author{R. Folman}
\affiliation{Department of Physics, Ben-Gurion University of the Negev, Be’er Sheva 84105, Israel}
\author{E. Cohen}
\affiliation{Faculty of Engineering and the Institute of Nanotechnology and Advanced Materials, Bar Ilan University, Ramat Gan 5290002, Israel}


\begin{abstract}
\textbf{Bohmian mechanics was designed to give rise to predictions identical to those derived by standard quantum mechanics, while invoking a specific interpretation of it -- one which allows the classical notion of a particle to be maintained alongside a guiding wave. For this, the Bohmian model makes use of a unique \emph{quantum potential} which governs the trajectory of the particle. In this work we show that this interpretation of quantum theory naturally leads to the derivation of interesting new phenomena. Specifically, we demonstrate how the fundamental Casimir-Polder force, by which atoms are attracted to a surface, may be temporarily suppressed by utilizing a specially designed quantum potential. We show that when harnessing the quantum potential via a suitable atomic wavepacket engineering, the absorption by the surface can be dramatically reduced. This is proven both analytically and numerically. Finally, an experimental scheme is proposed for achieving the required shape for the atomic wavepacket. All these may enable new insights into Bohmian mechanics as well as new applications to metrology and sensing.}

\end{abstract}

\maketitle



\section{Introduction}

Quantum mechanics (QM) challenges our common sense. For example, it allows
superposition states which we never see directly, it adheres to a minimal uncertainty principle, and it is non-local. This has brought its own founding fathers, such as Schr\"{o}dinger,
Einstein and de Broglie, to speak against it. This has also given rise to many
attempts to reinterpret it or even extend it. One of the attempts to reinterpret QM, and perhaps provide a base for
future extensions of the theory, has been initially developed by de Broglie
and Bohm, and has been termed Bohmian mechanics (BM)
\cite{Bohm,BohmHiley,Holland,Cushing}.
\begin{figure}[!t]
\includegraphics*[width=8cm]{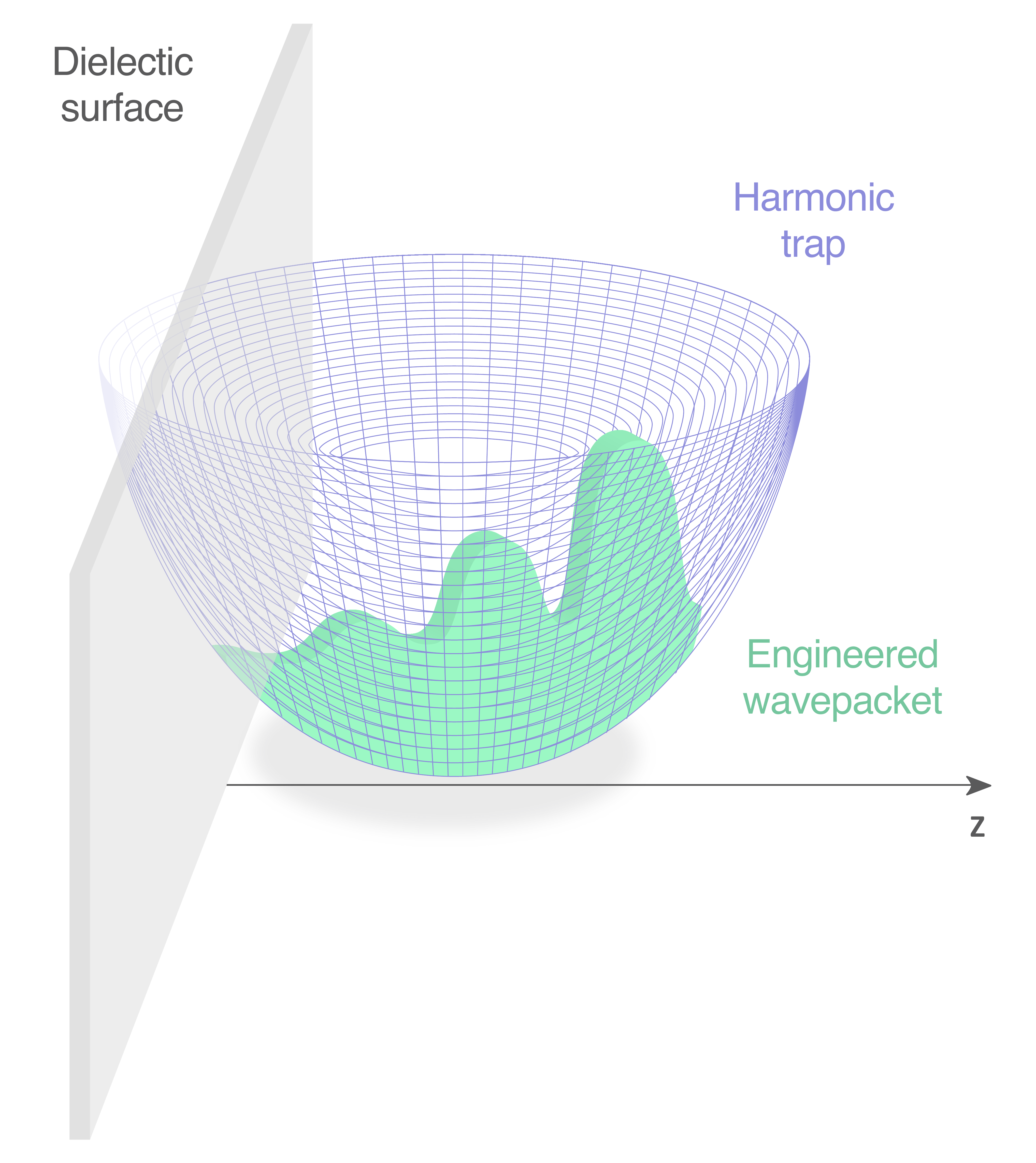}\caption{\textbf{A schematic
illustration of the setup}. A wavepacket trapped in a harmonic potential is brought (together with the trap) close to a dielectric surface. The specific $z$-dependence of the wavepacket is given by
Eq.\,\ref{tsolution}, and is depicted in Fig.\,2.}%
\label{figsetup}%
\end{figure}
Over the last decade its proponents have shown it to have a relativistic form
\cite{Durr99,Durr14}, as well as a quantum field theory form \cite{BMQFT}.
Recently, several experimental studies have reported the observation of  Bohmian trajectories
\cite{Wiseman,Stein1,Stein2}, but some have named these trajectories
\textquotedblleft surrealistic\textquotedblright\cite{Surr1,Surr2}.
Interestingly, several works have pointed out that using BM could help solving
complex numerical problems in QM \cite{Solv1,Solv2,Solv3}. However, the usefulness of this intriguing interpretation has remained under debate, and there does not seem to be a consensus yet as to the conceptual and practical merits of BM.

In this letter we show that the Bohmian quantum potential (defined in the next section) enables the engineering of new phenomena. Specifically we show how a fundamental force, the Casimir-Polder (CP)
force, may be suppressed using this unique potential. This may enable new insights into the foundations of quantum theory, and may allow for new pathways in quantum technology applications, such as metrology and sensing.
In particular, during the last two decades we have witnessed a major growth of experiments
with cold atoms near surfaces. These experiments were driven by the desire to increase integration and scalability while miniaturizing these promising quantum devices, e.g. for various applications in metrology and atom
interferometry \cite{Review1,Review2}. The CP
potential becomes important close to the surface, posing
both fundamental and practical challenges. Recent works have utilized cold atoms to study the CP potential and examine
atom-surface interactions \cite{Lin,Ketterle,Obr,Dalvit,Z1}.

We utilize the quantum potential $Q$, which depends only on the shape of the wavefunction, to propose a special engineering of atomic wavepackets which enables
them to partially resist the CP attraction to a nearby surface.

Within the next two sections we analyze, first analytically and then numerically, the proposed wavepacket engineering and its performance. We then briefly outline an experimental protocol for realizing this particular wavepacket.

\begin{figure*}[!ht]
\includegraphics*[width=\textwidth]{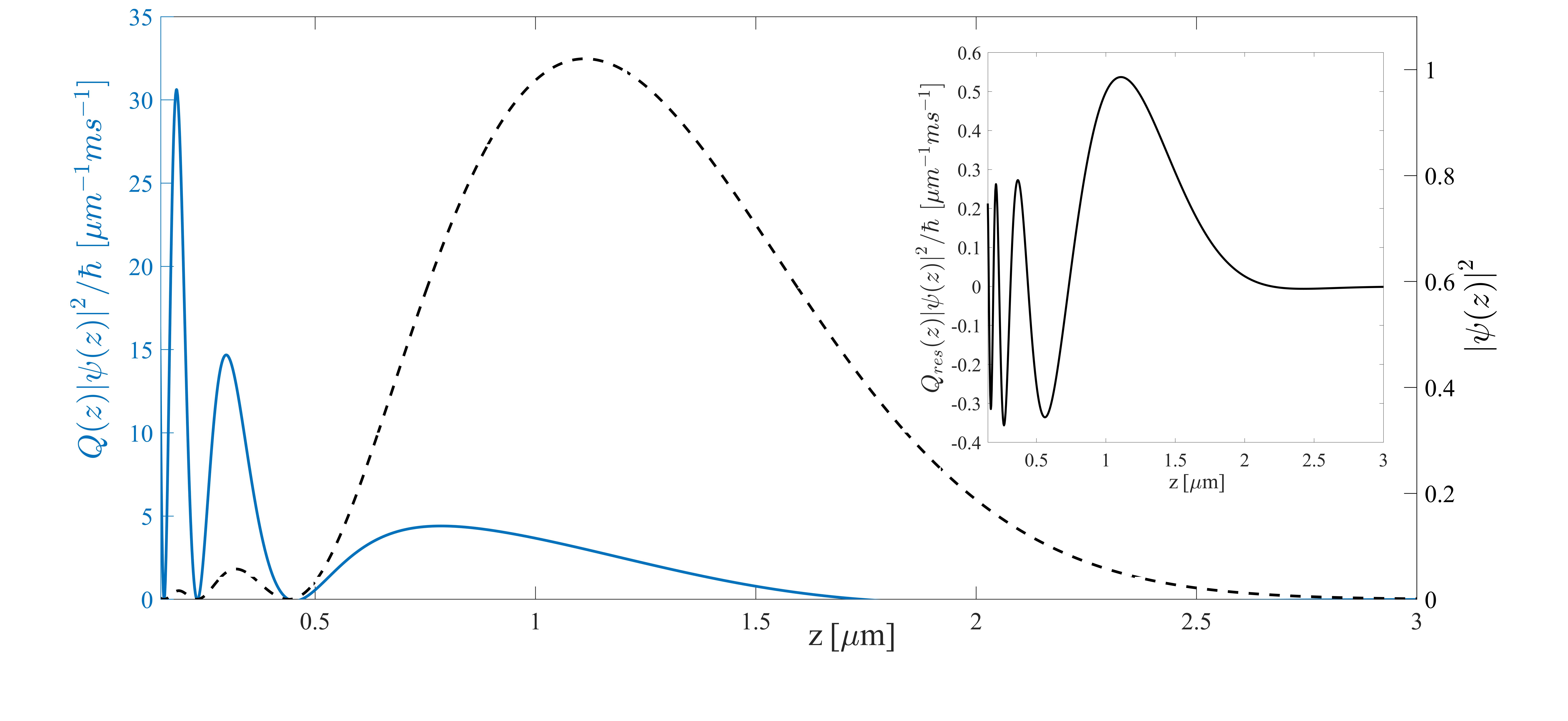}\caption{{\bf The engineered wavepacket and its quantum potential.} The dashed curve describes the density of the engineered wavepacket. The solid curve stands for the weighted quantum potential, namely, the quantum potential multiplied by the density of the engineered wavepacket (divided here by $\hbar$). The inset presents the weighted residual potential, namely, the quantum potential together with the Casimir-Polder potential, multiplied by the density of the engineered wavepacket (same units). The negligible residue shows that indeed the quantum potential is able to successfully counter the Casimir-Polder potential.}
\label{figQPotential}%
\end{figure*}

\section{Suppressing the CP force via atomic wavepacket engineering}


In what follows, we propose a theoretical method to effectively suppress the CP force (for a limited amount of time) through the generation of a tailored quantum potential. This
method can be simply described when applying the Madelung transformation
\cite{Mad} following the recent analyses in \cite{Madelung,ShortT}. BM, as
well as the Madelung formalism, allow to efficiently describe the interplay
between external potentials and the quantum potential and hence we find them
very suitable in this case, where we try to counteract the former potentials.

We shall represent the wavefunction $\psi(\mathbf{r})$ in the polar form
\begin{equation}
\Psi(\mathbf{r},t) = \sqrt{\rho}(\mathbf{r},t)e^{iS(\mathbf{r},t)/\hbar},
\end{equation}
where $\rho$ and $S$ are the density and phase, respectively, and use the well-known guiding equation for the velocity
\begin{equation}
\label{deB}\mathbf{u} = \nabla{\tilde S},
\end{equation}
where the tilde superscript represents quantities per
unit mass $m$, so that ${\tilde S} = {\frac{S}{m}}$. The real part of the
Schr\"{o}dinger equation then becomes the continuity equation
\begin{equation}
{\frac{D}{Dt}}\ln\rho= - \nabla\cdot\mathbf{u}\, ,
\end{equation}
where ${\frac{D}{Dt}} \equiv{\frac{\partial}{\partial t}} + \mathbf{u}%
\cdot\nabla$, is the material (Lagrangian) time derivative of a fluid element
along its trajectory, and the imaginary part becomes
\begin{equation}
{\frac{\partial\tilde S}{\partial t}} = -({\tilde K} + {\tilde Q} + {\tilde
U}),
\end{equation}
where ${\tilde K} = \mathbf{u}^{2}/2$ is the kinetic energy per unit mass and
\begin{equation}
\tilde{Q} = - {\frac{\hbar^{2} }{2m^{2}}}{\frac{\nabla^{2} \sqrt{\rho} }%
{\sqrt{\rho}}}%
\end{equation}
is the quantum potential per unit mass.

For an irrotational potential flow in the form of Eq.\,\ref{deB}, we then obtain
\begin{equation}
\label{2ndlaw}{\frac{D}{Dt}}\mathbf{u} = -\nabla\tilde{Q}(\rho) -\nabla{\tilde
U},
\end{equation}
suggesting the possibility of cancelling an external potential
$\tilde U$ using a suitable quantum potential $\tilde{Q}$.

Our proposed experimental setup consists of an atomic wavepacket $\psi(x,z,t)$.  At time $t=t_{0}$ the wavepacket is brought close to the vicinity of a planar dielectric surface situated at $z=0$ using a harmonic trap (see
Fig.\,\ref{figsetup}). We assumed that the harmonic trap remained there also for $t>0$, but assuming otherwise barely changed the simulated dynamics.

A CP potential $U(\mathbf{r})=-C_{4}/z^{4}$ acts
on the atoms close to the dielectric surface, where $C_{4}$ is some constant
depending on the properties of the surface and the atoms. Hereinafter we assume that the surface is the $z=0$ plane. We further assume that the wavepacket stays around the sub-micrometer distance from the surface but almost vanishes at the very close regime of about 100\,nm from the surface, where the atom-surface interaction is dominated by the van der Waals potential, which has a $z^{-3}$ dependence.
We can therefore arrange the
desirable situation ${\frac{D}{Dt}}\mathbf{u}=0$, where the total acceleration
of the atoms is zero, by preparing a density $\rho(\mathbf{r})\equiv
P^{2}(\mathbf{r})$, which satisfies
\begin{equation}
\label{condition}\nabla^{2} P + \frac{2mC_{4}}{\hbar^{2}z^{4}}P=0.
\end{equation}

In our proposed experimental setup we are only interested in the dynamics along the $z$ axis and hence our problem becomes one-dimensional (1D). In 1D, the solution of the ordinary differential equation corresponding to
Eq.\,\ref{condition} is%

\begin{equation}
\label{solution}P(z)=z\left[  C_{1}\cos\left(  \frac{\sqrt{2mC_{4}}%
}{z\hbar}\right)  +C_{2}\sin\left(  \frac{\sqrt{2mC_{4}}}{z\hbar}\right)
\right]  .
\end{equation}

See the Supplementary Material for additional details. We note that this wavefunction is continuous at $z=0
$, i.e. on the surface, and vanishes there. This function is not always positive but the physically meaningful field $\rho$ is. Our numerical simulations below indicate a slightly inferior performance of $|P(z)|$ compared to $P(z)$ and thus we use it hereinafter.



%


We now have to properly truncate the wavefunction for making it realistic
(and square-integrable). This can be done, for instance, by multiplying it
with a Gaussian envelope, thus reaching a wavepacket of the form
\small{
\begin{equation}
\label{tsolution}\psi(z)=ze^{-\frac{(z-z_0)^{2}}{4\sigma^{2}}}\left[  C_{1}\cos\left(
\frac{\sqrt{2mC_{4}}}{z\hbar}\right)  +C_{2}\sin\left(  \frac{\sqrt
{2mC_{4}}}{z\hbar}\right)  \right]  ,
\end{equation}}
where the constants $z_0$ and $\sigma$ and are the Gaussian's mean and width, respectively.
A wavefunction having this density will spread with time, but
as was shown in \cite{ShortT}, $\tilde{Q}(\mathbf{r},t+\Delta t)=\tilde
{Q}(\mathbf{r},t)+O((\Delta t)^{2})$. Therefore, an initial preparation of a
wavefunction according to Eq.\,\ref{tsolution} is a good estimation for short
times, which is the regime we will numerically simulate below. For larger times, the wavepacket  further spreads and becomes more and more distorted, thereby creating a different quantum potential which might be less beneficial.
Using this technique a suitably prepared atomic wavepacket can be used, e.g., for measuring magnetic fields near the surface while passing above it, without being strongly drawn towards it.

For the short time spent by the wavepacket in the vicinity of the surface, this kind of
truncation was shown in a similar context to be almost innocuous \cite{Siv}. In
our case, it hinders the cancellation of the CP potential as
calculated in the Supplementary Material, but not fatally. The Gaussian truncation results in an unwanted {\it residual potential} (hereinafter there is no division by $m$ and hence no tildes are used) which is equal to

\begin{equation} \label{EqQres} Q_{\rm res}=-\frac{\hbar^2}{2m\sigma^2}\left\{1+\frac{z-z_0}{\sigma}\left[2\sigma \frac{P'(z)}{P(z)}-\frac{z-z_0}{\sigma}\right]\right\}
\end{equation}

Although the residual potential is not negligible, its largest component near
$z=0$ scales like $1/z^{2}$, hence it suggests a major improvement in
comparison to the CP potential which scales like $1/z^{4}$ close to the
surface. Fig.\,\ref{figQPotential} shows the weighted quantum potential, the weighted residual potential and the density of the engineered wavepacket.

Moreover, the inverse proportionality to $2\sigma^{2}$ guarantees
that by increasing the width of the Gaussian envelope we can further shrink
the overall size of the residual quantum potential. The term $P'(z)/P(z)$ is also diverging, but if we average over the region of each singularity we will get a small contribution (sometimes in the form of a favorable repulsive potential). Therefore, and in contrast to the unengineered Gaussian, where the atoms are strongly attracted to the surface, here they will not be attracted so strongly. On the other hand, they may suffer from irregularities in the vicinity of the singularities and this is the reason that the numerical simulation performed below is important. Nevertheless, we may conclude on analytic
grounds that while a solution of the form Eq.\,\ref{solution} could completely
cancel the CP potential, the more realistic truncated shape in Eq.\,\ref{tsolution} also has the ability to suppress the CP force (for a limited amount of time).

\section{Numerical simulation}


In order to study the performance of the proposed wavepacket engineering, we perform a numerical simulation examining the dynamics of the wavepacket in the vicinity of a dielectric surface. In particular, we wish to examine how the absorption of particles evolves in time when using engineered and unengineered wavepackets. The aim, of course, is to minimize the absorption when using an engineered wavepacket.

Below we briefly outline the details of the simulation and then
compare the absorbed fraction achieved over time with our engineered atomic wavepacket
(Eq.\,\ref{tsolution}) to that of a Gaussian wavepacket.

To embed the proposed scheme within a more realistic setup, we are assuming that the atoms lie in a harmonic trap, situated close to the
surface (throughout this work, the frequency of the harmonic trap is standardly determined by $\sigma$ and $m$). The atoms are initially at rest. In addition, to simulate the absorption in the surface we assume, similarly to \cite{Scott}, an imaginary
potential growing linearly from zero at $z=\delta=0.15\,\mu$m till $z=0$ (implying that any atom that enters the region below $\delta =0.15\,\mu$m is absorbed within a certain time, unless having a sufficiently large velocity in the other direction), while the real part of the potential remains constant throughout $0<z<0.15\,\mu$m (we do require continuity in $z=0.15\,\mu$m).

This necessary modification of the total potential deteriorates the performance of
our engineered wavepacket, which was not originally meant to resist it. However, as we shall
show below, the simulative rate of absorption exhibited by the engineered
wavepacket was still lower than that of the customary Gaussian wavepacket, which is consistent with the fact that the residual potential is much smaller than the CP potential near the surface.

We consider a $^{87}$Rb atom (mass $m=1.44\cdot 10^{-25}$\,kg) near a silicon surface (refractive index $n=2$). For the ground state static polarizability of $^{87}$Rb [$\alpha_0=0.0794$\,Hz/(V/c m)$^2$] we have $C_4=9.1\cdot 10^{-56}$\,J$\cdot$m$^4$. We now solve the time-dependent Schr\"odinger equation for the atoms under the influence of the Casimir Polder + Harmonic + absorbing potentials. We run a simulation for a large set of means ($z_{0}$) and standard deviations ($\sigma
$), and calculate the absorption fraction. Before the truncation, the sine and cosine solutions in
Eq.\,\ref{solution} should give rise to the same quantum potential. However, the residual potential stemming
from the truncation reveals that there is a difference between the two cases and we indeed notice some
advantage of the cosine solution over the sine.

\begin{figure}[tbp]
\includegraphics*[width=8cm]{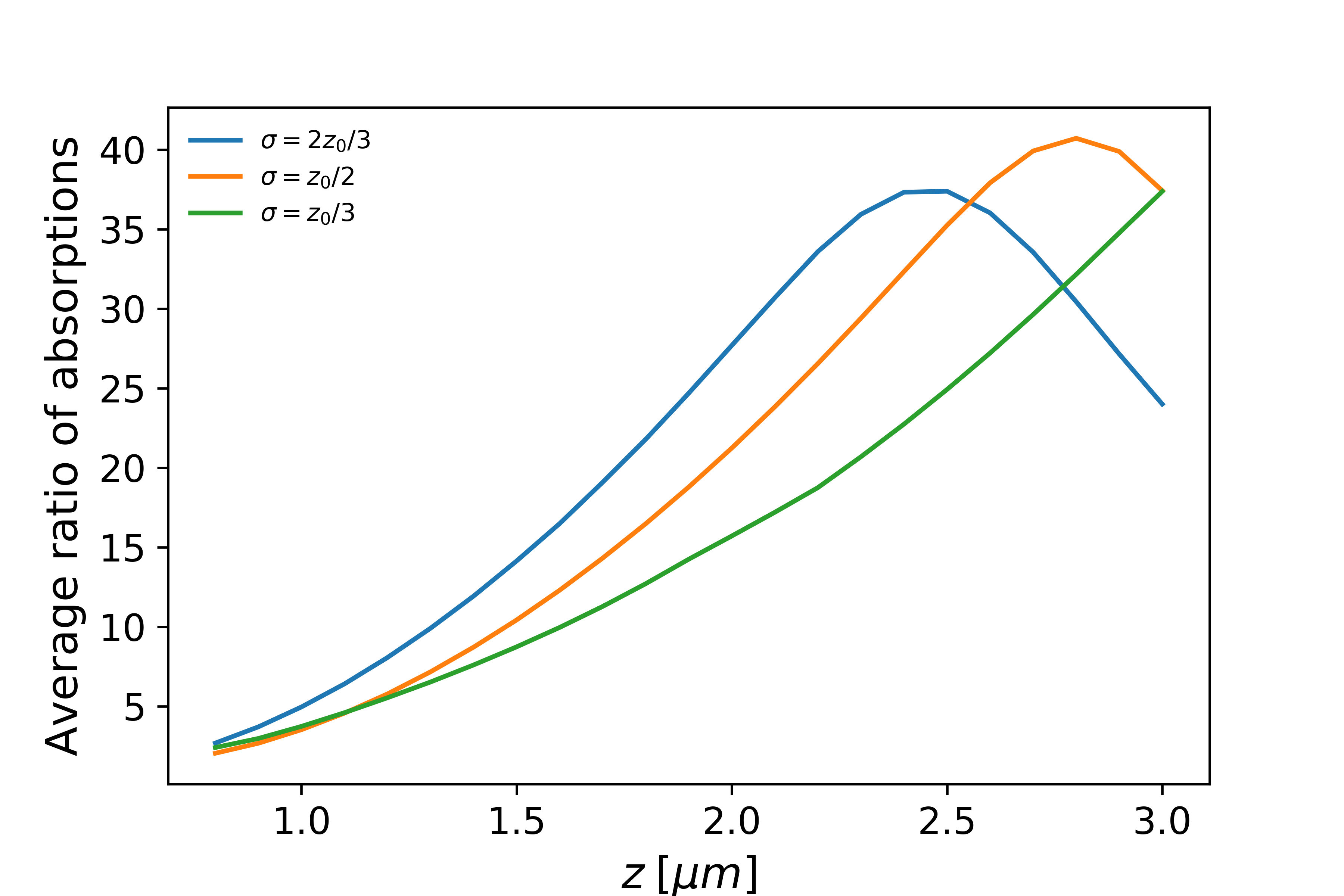}\caption{\textbf{Comparison, in terms of absorption, between the engineered wavepacket and a Gaussian for various preparations of the wavepackets.} The average ratio (over the time interval $0 \le t \le 2$\,ms) of the absorbed fractions is calculated (i.e. Gaussian wavepacket absorption divided by engineered wavepacket absorption) as a function of $z_0$, the mean position of the Gaussian envelope, while the ratio between the standard deviation and $z_0$ is kept constant (equal either to 2/3, 1/2 or 1/3). As expected, at short times large standard deviation is preferable, but in all cases a substantial advantage is achieved.}  \label{figcompz}.
\end{figure}

\begin{figure}[btph]
\includegraphics*[width=8cm]{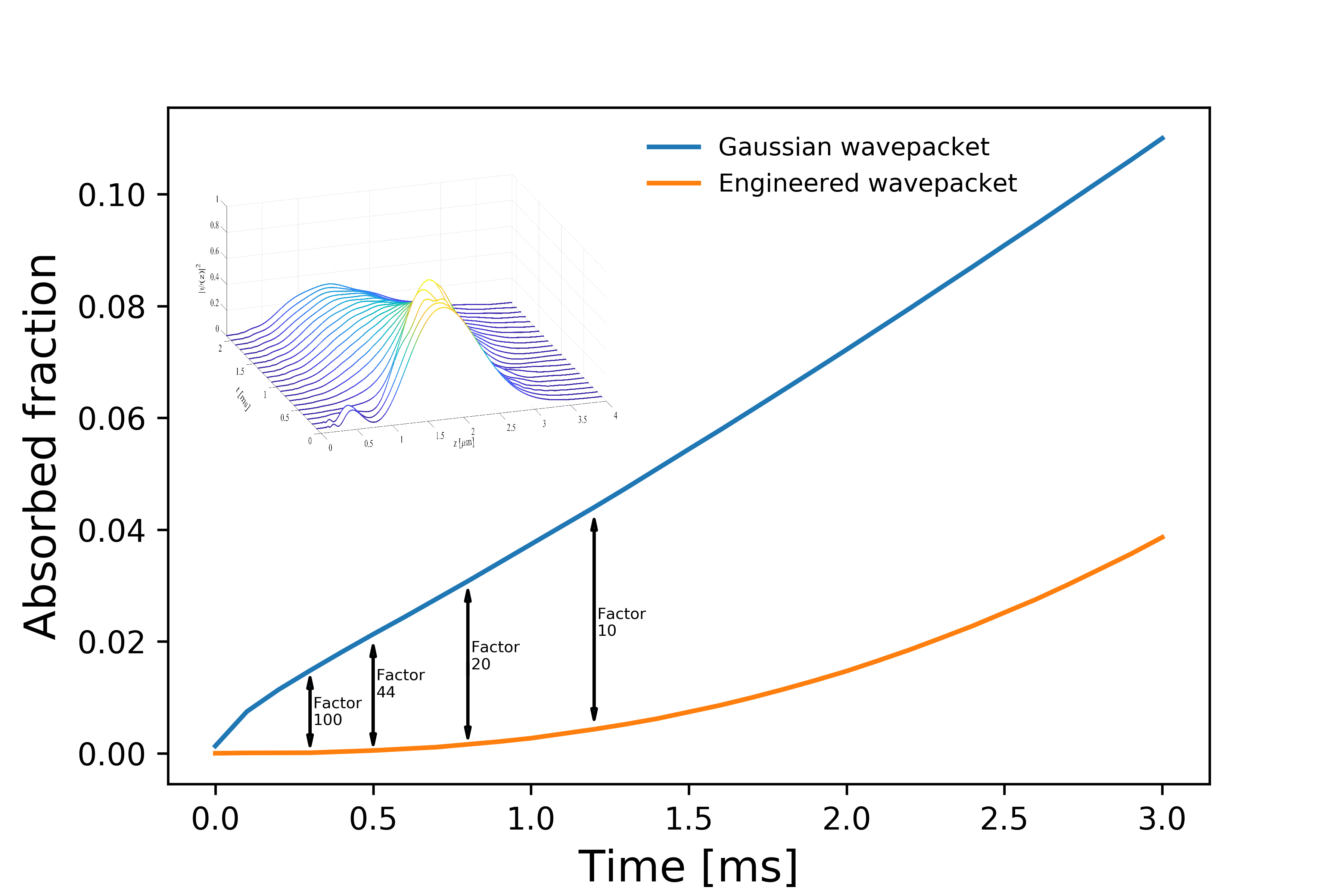}\caption{\textbf{Comparison between the engineered wavepacket and a Gaussian over time for a specific choice of parameters.} The mean and standard deviation of the Gaussian are chosen to be $z_0=3\,\mu$m and $\sigma=1\,\mu$m, respectively, enabling a particularly beneficial performance (two orders of magnitude advantage at short times and one order of magnitude at longer times). The absorption rates of the two wavepackets become comparable around 3\,ms. The inset shows the time evolution of the engineered wavepacket within the harmonic and Casimir-Polder potentials. As may be seen, the wavepacket eventually loses its specially engineered shape, thus diminishing the effect countering the Casimir-Polder force. For this specific calculation, the parameters used for the mean and standard deviation of the Gaussian envelope were particularly challenging, $z_0=1\,\mu$m $\sigma=\,2/3\,\mu$m, in order to emphasize the transient nature of the effect due to the external potentials.}\label{figcompt}.
\end{figure}

We compare for various cases our proposed solution in Eq.\,\ref{tsolution} to a Gaussian wavepacket with the same mean and standard deviation as those of the envelope -- See Fig.\,\ref{figcompz}. It is desirable, of course, to decrease as much as possible the residual potential, and since it contains three terms depending on $\sigma^{-2}$, $\sigma^{-3}$ and $\sigma^{-4}$, we have better results for higher-$\sigma$ wavepackets. To further explore the advantage we choose some specific parameters ($z_0=2.3\,\mu$m and $\sigma=1\,\mu$m) and compare the absorbed fractions -- See Fig.\,\ref{figcompt}.

As can be seen from Figs.\,\ref{figcompz} and \ref{figcompt}, the engineered wavepacket leads to a
significantly smaller absorbed fraction, which implies that the CP force has
less impact on our engineered wavepacket in comparison to its impact on a Gaussian wavepacket.
Thus, the results show that the engineered wavepacket's shape indeed has the
ability to reduce the unwanted effects of the CP force.
In accordance with our analytic expectations, the engineered wavepacket excels at
high standard deviations and short times, reaching in some cases a 100-fold advantage over the Gaussian wavepackets in terms of absorption.

Note, however, that the engineered wavepacket is typically skewed away from the surface in comparison to the Gaussian envelope (and hence in comparison to the Gaussian wavepacket we used as a benchmark). Thus, one may ask whether the presented advantage follows only from this spatial displacement rather than the special shape of our wavepacket. To test this hypothesis we tried to fit a Gaussian wavepacket to the engineered solution by locating it farther away from the surface and/or shrinking its width until it highly resembled the Gaussian we compared it to. In all these cases we still found an advantage in favor of the engineered wavepacket (albeit smaller). We present such a comparison in the Supplementary Material,  where we significantly pushed the Gaussian away from the surface. We still found that the engineered wavepacket has a substantially smaller absorbed fraction. Thus, we conclude that not only the shifted mean and modified width, but also the particular shape of the wavepacket contributed to the observed advantage.

Finally, we discuss the transient nature of the effect. As can be seen in Fig.\,\ref{figcompt}, the absorption rate of the engineered wavepacket shows no advantage after about 3\,ms. This is due to the fact that with time it loses its unique shape which originally gave rise to the required quantum potential. To show this explicitly, we plot in the inset of Fig.\,\ref{figcompt} the time evolution of the engineered solution from Eq.\,\ref{tsolution} with $z_0=1\,\mu$m and $\sigma=2/3\,\mu$m. We can clearly see the amplitude decreasing with time and the shape being distorted.

\section{A possible scheme for achieving the required wavepacket engineering}


As was shown in the previous
section, a certain shape of the wavepacket provides the
desirable result, i.e. resisting the CP force for a short time.
The construction of such a wavepacket can be obtained in
several ways. We discuss here in general terms a technique which utilizes external potentials and fields (resembling \cite{Ron's}).

First, a Gaussian atomic wavepacket can be prepared by cooling the atoms to occupy the Gaussian ground state of a harmonic potential. Engineering the wavepacket to the form of Eq.\,\ref{tsolution} may be done in two stages by utilizing an interferometric sequence  with two internal atomic states having a different response to an external potential, e.g., due to a magnetic field. We use a magnetic field gradient pulse of duration $T$ to create a state-dependent spacially varying potential sandwiched between two properly designed Rabi pulses inducing transitions between the two states. These would be followed by a projection to one state, transforming the atomic wavepacket as $\psi(z)\to \psi(z)[ae^{i\varphi(z)}+be^{-i\varphi(z)}]$. Here $\varphi(z)=\delta V(z)T/2\hbar$ is the phase imprinted by the short pulse of potential difference $\delta V(z)$ between the two atomic states, while $a$ and $b$ are complex numbers determined by the Rabi pulses (we neglect a possible space-dependent global phase).
In order to generate the linear $z$-dependence in front of the right-hand-side of Eq.\,\ref{tsolution} we can apply a linear potential difference $\delta V(z)= Fz$, choose $b=-a$, and obtain $\psi(z)\to \psi(z)\sin(Kz)\approx  Kz\psi(z)$ (if
$Kz=FTz/2\hbar\ll 1$). For generating the cosine (or sine) dependence with an argument proportional to $1/z$ we can apply a differential potential with a $1/z$ dependence, e.g. a magnetic field generated by a wire on the surface, and choose $|a|=|b|$ with a certain phase difference that determines the phase of the cosine function.

The resulting wavepacket, serving as a good approximation to Eq. \,\ref{tsolution} (for
$z>0$), could now be used e.g. as an input for a magnetometry experiment near the
dielectric surface without being strongly attracted to it.

\section{Discussion}


In summary, we explored the possibility to use the Bohmian potential in
order to cancel external potentials via wavepacket shaping. We have analytically proposed a novel technique that allows to suppress the
CP force (for a limited period of time) and tested in numerically. Using this approach, we examined the
case in which a Gaussian wavepacket is engineered into a special form such
that near a dielectric surface the wavepacket resists the CP force. We have addressed the case of a wavepacket brought to a surface using a harmonic trap, but the presented analysis can be readily generalized to a grazing beam scenario (such as \cite{Z1}). Although being related in the past to {\it surreal} phenomena, this
work emphasizes the very {\it real} effects of the quantum potential, as well as
its possible applications in practical scenarios. In addition to providing new insight concerning the Bohmian interpretation, this analysis may pave the way for various applications, e.g. surface magnetometry with cold atoms having engineered wavepackets which can survive longer at the vicinity of the surface. Although the CP potential was analyzed above, it should be noted that the useful interplay between external potentials and the quantum potential is general. Therefore, by carefully engineering the wavepacket, it is possible to suppress additional forces such as the gravitational or van der Waals forces.


\section{Methods}


\subsection{The engineered wavepacket and the total potential}

The engineered wavepacket we employed is given by the solution of

\begin{equation}
\label{condition}\nabla^{2} P + \frac{2mC_{4}}{\hbar^{2}z^{4}}P=0,
\end{equation}

multiplied by a Gaussian truncation $e^{-\frac{(z-z_0)^{2}}{4\sigma^{2}}}$ , yielding

\begin{equation}
\label{tsolution}\psi(z)=ze^{-\frac{(z-z_0)^{2}}{4\sigma^{2}}}\left[  C_{1}\cos\left(
\frac{\sqrt{2mC_{4}}}{z\hbar}\right)  +C_{2}\sin\left(  \frac{\sqrt
{2mC_{4}}}{z\hbar}\right)  \right].
\end{equation}




In our numerical simulation we assumed that the wavepacket is brought to the vicinity of the dielectric surface using a harmonic potential which remains there. Then, following \cite{Scott}, the Casimir-Polder potential $V_{CP}(z)$ is acting on the wavepacket for any $z \ge 0.15\,\mu$m. For $0<z<0.15\,\mu$m it is replaced by a constant value, $V_{CP}(0.15)$ (assuring continuity at $z = 0.15\,\mu$m), and a linear imaginary potential is added as in \cite{Scott}.


\subsection{Calculation of the residual potential}

We shall compute here the residual quantum potential corresponding to our proposed construction. This will allow us to investigate more deeply the analytic properties of the shaping, which are important for any experimental demonstration of this technique.

In 1D, the solution of our ordinary differential equation takes the form
(\ref{tsolution}). For simplicity we define the following function%
\begin{equation}
\alpha\left(  z\right)  =\left[  C_{1}\cos\left(  \frac{\sqrt{2mC_{4}}%
}{z\hbar}\right)  +C_{2}\sin\left(  \frac{\sqrt{2mC_{4}}}{z\hbar}\right)
\right]  .
\end{equation}
We shall now compute the value of
\begin{equation}
\chi=\frac{d^{2}}{dz^{2}}\left(
e^{-\left(  z-z_0\right)  ^{2}/\zeta}z\alpha\left(  z\right)  \right)
+\frac{2mC_{4}}{\hbar^{2}z^{4}}\left(  e^{-\left(  z-z_0\right)  ^{2}/\zeta
}z\alpha\left(  z\right)  \right),
\end{equation}
where $\zeta=8\sigma^{2}$. That is%
\begin{equation}
\chi=\frac{d}{dz}\left(  \frac{d}{dz}ze^{-\left(  z-z_0\right)  ^{2}/\zeta
}\alpha\left(  z\right)  \right)  +\frac{2mC_{4}}{\hbar^{2}z^{4}%
}ze^{-\left(  z-z_0\right)  ^{2}/\zeta}\alpha\left(  z\right)  ,
\end{equation}

and after some algebraic calculations%
\begin{equation}
\begin{array}{lcl}
\chi =e^{-\left(  z-z_0\right)  ^{2}/\zeta}[-\frac{2}{\zeta}
3z\alpha\left(  z\right)  -2z_0\alpha\left(  z\right)  +z^{2}%
\alpha^{\prime}\left(  z\right)  -zz_0\alpha^{\prime}\left(  z\right)
\\+z\left(  z-z_0\right)  \alpha^{\prime}\left(  z\right)
+\frac{4}{\zeta^{2}}z\left(  z-z_0\right)  ^{2}\alpha\left(  z\right)
+z\alpha^{\prime\prime}\left(  z\right)  +\frac{2mC_{4}}{\hbar^{2}z^{3}%
}\alpha\left(  z\right) \\ +2\alpha^{\prime}\left(  z\right)  ].
\end{array}
\end{equation}

From the fact that $\frac{2mC_{4}}{\hbar^{2}z^{3}}\alpha\left(  z\right)
+2\alpha^{\prime}\left(  z\right)  +z\alpha^{\prime\prime}\left(
z\right)  =0,$ we have%
\begin{equation}
\begin{array}{lcl}
\chi=\frac{e^{-\left(  z-z_0\right)  ^{2}/\zeta}}{\zeta}[\left(
-6z+4z_0\right)  \alpha\left(  z\right)  -4z\left(  z-z_0\right)
\alpha^{\prime}\left(  z\right)  \\+\frac{4}{\zeta}z\left(  z-z_0\right)
^{2}\alpha\left(  z\right)  ].
\end{array}
\end{equation}

This suggests that the residual quantum potential is
\[
Q_{res}=\frac{\hbar^{2}}{2m}\left[  \frac{-6+4z_0/z-4(z-z_0)\alpha%
^{\prime}\left(  z \right)  /\alpha\left(  z\right)  }{\zeta}%
+\frac{4\left(  z-z_0\right)  ^{2}}{\zeta^{2}}\right]  .
\]
This can be written more shortly as
\[ Q_{\rm res}=-\frac{\hbar^2}{2m\sigma^2}\left\{1+\frac{z-z_0}{\sigma}\left[2\sigma \frac{P'(z)}{P(z)}-\frac{z-z_0}{\sigma}\right]\right\} \label{ResPot} \]
where  $P(z)=z\alpha(z)$ is the original function defined in Eq. \,\ref{solution} of the main text. This coincides with Eq.\,\ref{EqQres}.

\subsection{Comparison to a fitted Gaussian}

\begin{figure}[h]
\centerline{
\includegraphics*[width=9cm]{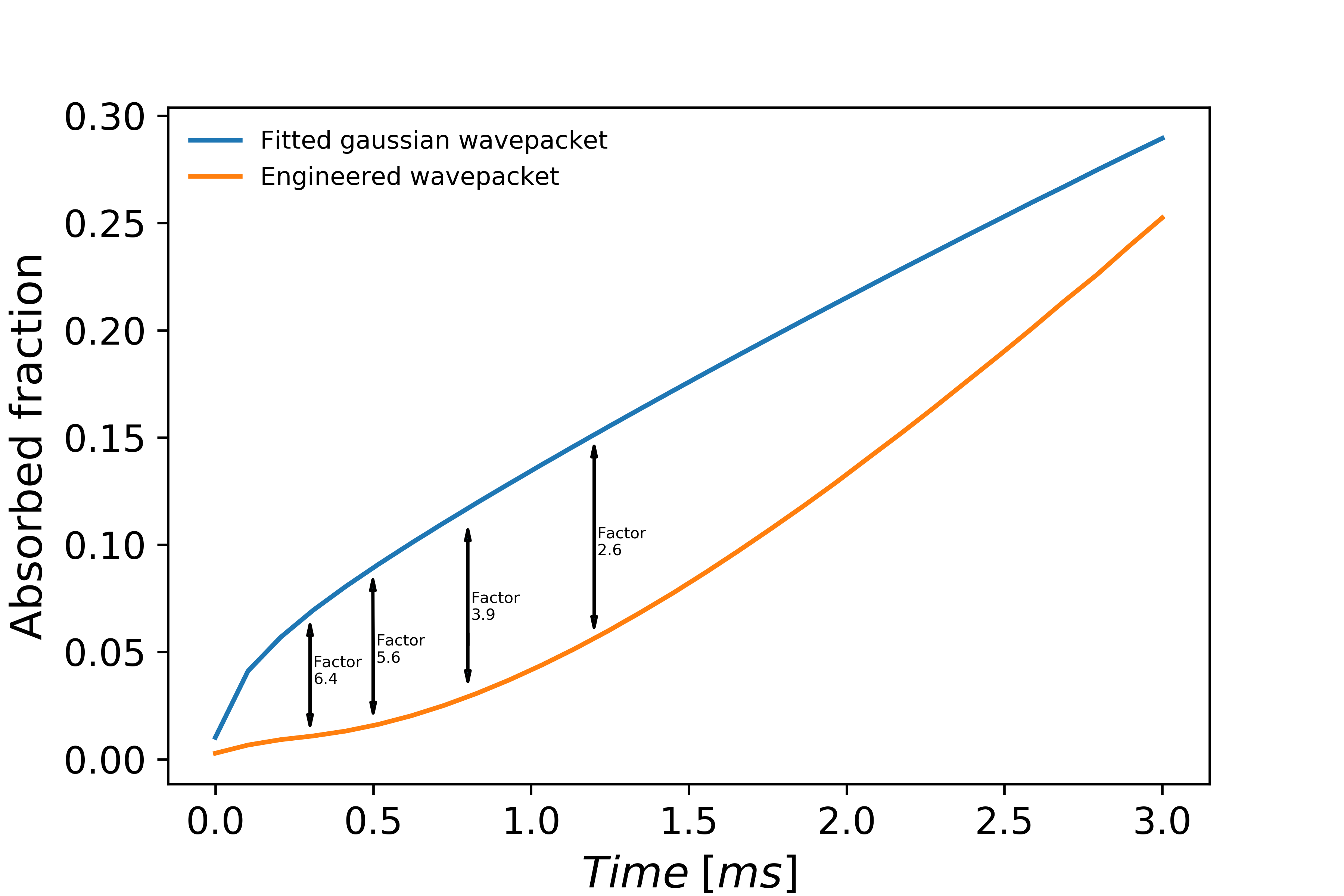}}\caption{\textbf{Comparison between an engineered wavepacket and a fitted Gaussian over time for a specific choice of parameters.} The mean and standard deviation of the Gaussian are chosen to be $z_0=2.3\,\mu$m and $\sigma=1\,\mu$m, respectively, while the corresponding parameters of the engineered wavepacket are $z_0=1.43\,\mu$m and $\sigma=1\,\mu$m. Despite the significantly improved parameters of the Gaussian, the engineered wavepacket maintains a substantial advantage. }\label{wcomp}
\end{figure}

To verify that the advantage reported in the main text, in terms of absorption, stems from the shape and not only the bias of the engineered wavepacket away from the surface, we compare in Fig.\,\ref{wcomp} the absorption of the engineered solution to the absorption of a fitted Gaussian located at the same distance from the dielectric surface. The advantage is smaller, but still apparent.

\end{document}